\documentclass[sigconf,author]{acmart}

\AtBeginDocument{%
  }

\usepackage[table]{xcolor}
\usepackage{multirow}

\usepackage{algorithm}
\usepackage{algpseudocode}

\begin{document}

\title[DRC-Aid]{DRC-Aid: \underline{D}esign-\underline{R}ule \underline{C}orrection via \underline{A}gentic Framework utilizing \underline{I}nference-Time Large Language Mo\underline{d}els }

\author{Anushka Mukherjee}
\affiliation{%
  \institution{Purdue University}
  \city{West Lafayette}
  \state{Indiana}
  \country{USA}}

\author{Kang He}
\affiliation{%
  \institution{Purdue University}
  \city{West Lafayette}
  \state{Indiana}
  \country{USA}
}

\author{Kaushik Roy}
\affiliation{%
  \institution{Purdue University}
  \city{West Lafayette}
  \state{Indiana}
  \country{USA}
}

\begin{abstract}

Resolving Design Rule Violations (DRVs) in layouts entails an iterative loop of geometric edits and verification. We present \textbf{DRC-Aid}, a closed-loop agentic framework that automates local DRC repair by formulating it as verification-in-the-loop search. To constrain the combinatorial geometric repair space, a deterministic Rule Engine converts physical verification tool-reported violations into a bounded menu of geometric edits. An off-the-shelf Large Language Model (LLM) evaluates local geometric context to select edits from this menu, with budgeted depth-first search and backtracking. Immediate feedback from verification tools such as \textit{Calibre nmDRC/nmLVS} enforces geometric compliance and guards against electrical-topology degradation, while a global Memory Bank prevents cyclic re-exploration. Evaluated on FreePDK45 layouts containing DRVs, DRC-Aid achieves DRC-clean, LVS-equivalent repairs in $\sim$92.5\% of cases with a $\sim$98\% total violation reduction, while residual cases yield partially repaired LVS-equivalent candidates. Under an identical search and verification infrastructure, LLM-based selection outperforms random (54.4\%) and deterministic-heuristic (83.3\%) policies, with the gap widening on cases with six or more violations.

\end{abstract}



\settopmatter{printacmref=false}
\setcopyright{none}
\renewcommand\footnotetextcopyrightpermission[1]{}
\pagestyle{plain}

\maketitle

\section{Introduction}\label{sec:intro}
Physical verification ensures that Integrated Circuit (IC) layouts satisfy fabrication constraints. A critical component of this stage is Design Rule Checking (DRC), which verifies that layout geometry complies with the constraints imposed by the fabrication process~\cite{lavagnoIC}. During layout development, engineers routinely encounter numerous Design Rule Violations (DRVs), requiring repeated cycles of geometric modification and re-verification to reach a DRC-clean state~\cite{layout-to-generator,sar-adc-manual-design-analog}. A central challenge is context-dependent decision-making. Resolving one violation can displace neighboring polygons, introducing new violations that extend the design cycle.

Existing machine learning approaches primarily target DRC hotspot prediction and violation localization~\cite{PGR-DRC,CNN-PGR,CNN-DRCmap}; they identify where violations occur but offer limited guidance on correcting them. The automated repair approach of~\cite{Simm-Ann} relies on simulated-annealing optimization without adaptive reasoning. Large Language Models (LLMs) are a compelling alternative given their demonstrated trade-off reasoning~\cite{ChipNemo} and structured EDA tasks~\cite{RTLcoder,autochip,DRC-Coder}. However, directly applying LLMs to physical design remains error-prone: they hallucinate components and violate strict physical constraints~\cite{circuitLM, LLM-future-or-mirage}. Recent literature therefore argues that such systems must operate as agents that maintain state, plan action sequences, invoke deterministic tools, and adapt in a closed-loop manner~\cite{agentic-AI-pd}.

Building on this paradigm, we present \textbf{DRC-Aid}, a closed-loop agentic framework that shifts the LLM's role from an unconstrained edit generator to a highly constrained decision-maker. DRC-Aid formulates local layout repair as \emph{verification-in-the-loop search}: a deterministic \textbf{Rule Engine} converts each Calibre-reported violation into a bounded menu of executable geometric edits, an off-the-shelf LLM within the \textbf{Reasoning Agent (RAT)} selects an edit from among them, and the \textbf{Layout Management Toolkit (LMAT)} executes each edit and validates it via \textit{Calibre}. A \textbf{Mega Controller (MC)} orchestrates this loop through budgeted depth-first search (DFS) with backtracking, supported by a global memory of visited states. 

DRC-Aid targets the repetitive local correction loop encountered during transistor-level layout development, where an engineer identifies a geometric violation, applies a small modification, reruns DRC, and revises the edit when neighboring violations appear. We investigate whether this bounded but frequently repeated class of local corrections 
can be automated through a constrained, tool-grounded agentic workflow.

\begin{figure*}[t]
  \centering
  \includegraphics[width=\textwidth]{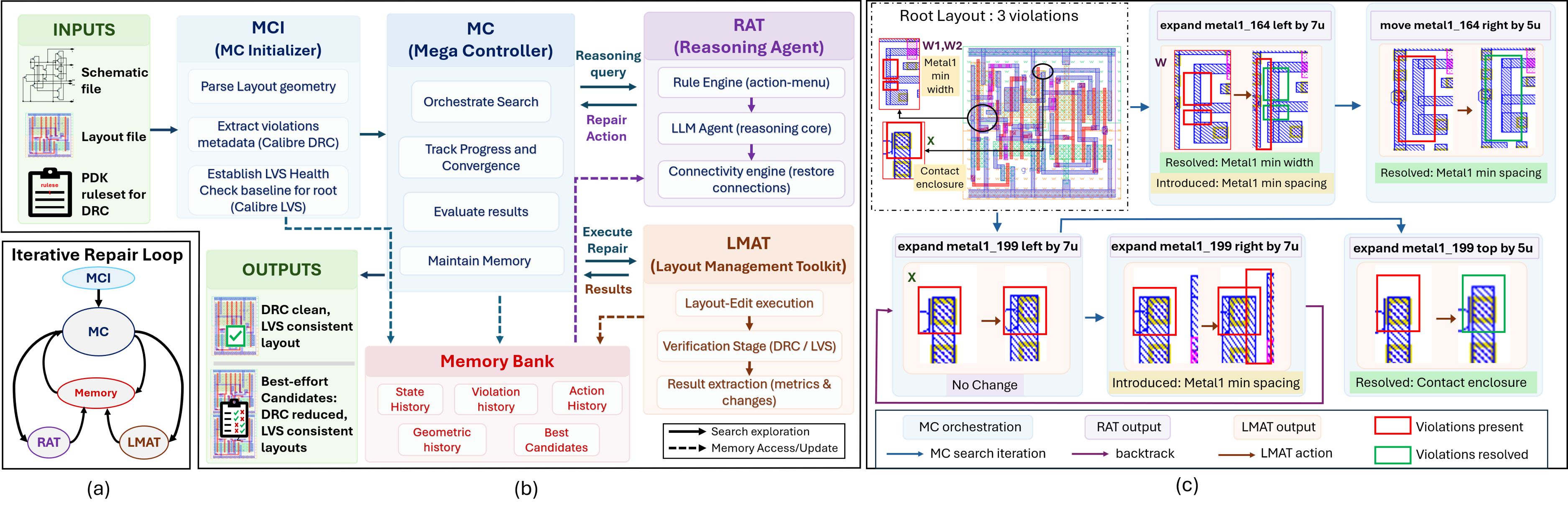}
  \caption{a) Iterative Search Exploration Loop of DRC-Aid b)
  Overview of the DRC-Aid Framework c) Illustrations of the iterative repair DRC-Aid conducts: the first row shows how fixing an initial violation caused another, which was then resolved; the second row shows that iterative steps induced a new violation instead of resolving the target; DRC-Aid backtracked and selected an alternative action that resolved it}
  \Description{Figure a shows the iterative loop that DRC-Aid goes through to reach a potential solution. Figure b provides the framework. It begins with the inputs, which then pass through the MCI to the MC, where search orchestration takes place. The RAT and LMAT are invoked to continue exploration until a solution in the form of DRC-clean state or Pareto condition of DRC-reduced state is achieved. Figure c illustrates the iterative repair DRC-Aid conducts: the first row shows how fixing an initial violation caused another, which was then resolved; the second row shows that iterative steps induced a new violation instead of resolving the target; DRC-Aid backtracked and selected an alternative action that resolved it}
  \label{fig:main1}
\end{figure*}

The key contributions of this work are:
\begin{itemize}
 \item \textbf{DRC-Aid}, an agentic framework that automates local geometric DRC repair by coupling off-the-shelf LLM reasoning with verification-guided search in a closed loop.
 \item A deterministic \textbf{Rule Engine} that discretizes the combinatorial geometric repair space into a bounded, executable action menu, enabling targeted LLM spatial reasoning.
 \item A \textbf{budgeted DFS} search with backtracking and a \textbf{global memory bank} that prunes suboptimal branches and prevents cyclic re-exploration.
\item Strict (DRC-clean, LVS-equivalent) repair in $\sim$92.5\% of cases with $\sim$98\% total violation reduction on FreePDK45~\cite{FreePDK} test cases. Under identical infrastructure, LLM action selection outperforms random (54.4\%) and heuristic (83.3\%) baselines.
\end{itemize}



\section{Background and Related Works} \label{sec:bg+rel_works}
\subsection{DRC in Physical Design}
Physical design transforms a circuit representation into a fabrication-ready layout through placement, routing, and physical verification stages \cite{kahng2011vlsi}. Verification primarily includes DRC to ensure geometric compliance and Layout-vs-Schematic (LVS) checks to confirm electrical equivalence \cite{lavagnoIC}. A DRV occurs when layout geometry fails to satisfy the rules specified by the fabrication process. Foundries define these rules within a Process Design Kit (PDK). Resolving DRVs is iterative: engineers use tools (e.g., \textit{Calibre nmDRC}) to extract violation markers, determine necessary modifications, apply geometric edits, and re-verify to ensure no new violations are introduced. Although rule definitions and thresholds vary across technologies, commonly encountered categories include minimum width, spacing, enclosure, and overlap.

\subsection{Existing Works}
Existing research in DRC automation bifurcates into two categories: violation prediction and stage-specific repair. Prediction-focused works like \cite{PGR-DRC} propose unsupervised methods to flag DRC hotspots pre-global routing. Similarly, CNN-based frameworks \cite{CNN-PGR,CNN-DRCmap} demonstrate that models can learn to predict violation density maps by extracting spatial features and congestion data. While these methods offer early identification of high-risk regions to guide global optimization, they are fundamentally diagnostic in nature: they identify \textit{where} the violations may occur but do not provide solutions to fix them. Existing repair strategies are generally restricted to specific design stages. For instance, \cite{TritonRoute} embeds DRC-aware loops for routing closure, while \cite{NVCell} utilizes Reinforcement Learning to fix routing DRCs within standard cells. Although \cite{Simm-Ann} uses an improved simulated annealing algorithm to repair violations in standard cell layouts, it treats repair as a cost-minimization problem using stochastic perturbations rather than context-aware spatial reasoning. Recent LLM-in-EDA works \cite{ChipNemo,LayoutCopilot,DRC-Coder} have demonstrated the efficacy of LLMs in handling hardware tasks. However, they lack a closed-loop path from report to verified repair. DRC-Aid bridges this gap by introducing a context-aware, reasoning-driven iterative process by treating verification not as a downstream check but as the signal that drives each repair step.

\section{Methodology} \label{sec:methodology}
DRC-Aid is an iterative, context-aware framework designed to achieve autonomous DRC correction in transistor-level layouts. Fig.~\ref{fig:main1}(b) presents the core components and overall functionality of the framework. DRC-Aid utilizes \textbf{RAT (Reasoning Agent)} to navigate the repair search space; the \textbf{LMAT (Layout Management Toolkit)} interfaces with commercial EDA tools (\textit{Cadence Virtuoso} and \textit{Siemens Calibre}) for physical realization. The RAT and LMAT operate in tandem, with \textbf{MC (Mega-Controller)} guiding exploration.
DRC-Aid operates in a 4-phase iterative loop: Observe, Reason, Execute, and Validate/Verify. In the initial \textbf{observation} phase, the MC’s initialization module (MCI) parses the layout geometry and performs a DRC check to identify violation data. This serialized context is injected into the RAT alongside a localized action-menu generated by the Rule Engine (refer to Sec.~\ref{subsubsec:rule-eng}). The LLM then \textbf{reasons} to select the most promising fix. Following this, the Connectivity Engine runs a local connectivity-check on the current node against the original root layout (refer to Sec.~\ref{subsubsec:conn-eng}) and, when needed, bundles a compensatory routing edit with the LLM-generated fix to maintain topological integrity. The LMAT \textbf{executes} the bundled fix on the layout and triggers a DRC \textbf{verification}. The resulting violation counts and modified geometry are returned to the MC, which updates the Memory Bank and proceeds to the \textbf{observation} phase of the next iteration. Fig.~\ref{fig:main1}(c) illustrates how DRC-Aid resolves violations through an iterative exploration. 

\subsection{Mega-Controller (MC)} \label{subsec:MC}
The Mega Controller orchestrates the layout repair search-space exploration by structuring the repair process as a Budgeted DFS strategy. MC guides the search towards a solution while preventing the RAT from stagnating in suboptimal states.

\subsubsection{\textbf{Budgeted DFS and anchor protection}} \label{info:budget,DFS,anchor}
To effectively explore the search space and prevent exhaustion of computational resources on unpromising branches of the search tree, MC enforces a depth-and-state-dependent budgeted exploration. Each node is assigned a budget inversely proportional to its depth from the root layout node, naturally decaying the exploration breadth as depth increases. However, if a newly generated node constitutes an improved state (decreased violation count over the root), MC marks it as \textbf{anchor-protected}. This status grants an additional budget (proportional to its improvement) to encourage extensive exploration of that promising branch. 

\subsubsection{\textbf{Backtracking}} \label{info:backtrack}
The MC utilizes backtracking to escape sear-ch stagnation. While the MC dictates the overall exploration strategy, it permits the RAT to explicitly request a backtrack to a previous node. Backtracking away from anchor-protected nodes is temporarily restricted to allow aggressive exploration around improving states. Additionally, MC initiates an independent backtrack to a previous node if a modified layout fails the mandated post-LMAT health-check (refer to Sec.~\ref{LVS}), effectively pruning invalid candidates regardless of DRC status.

\subsubsection{\textbf{Graceful Termination}}\label{pareto}
If a fully DRC-Clean and electrically equivalent state is unattainable, MC prevents complete failure by performing a global state evaluation across the search tree. During exploration, the memory bank (refer to Sec.~\ref{subsec:memory}) maintains a dynamic subset of \textbf{best-so-far} nodes that exhibit minimal violation count. Upon termination (budget exhaustion or a RAT stop request), MC performs topological health-checks (refer to Sec.~\ref{LVS}) on these candidates. The framework then concludes by returning a list of \textbf{best-effort} solutions. This provides the physical design engineer with an improved starting point to resolve residual violations.

\subsection{Reasoning Agent (RAT)} \label{subsec:RAT}
\begin{figure}[ht]
  \centering
  \includegraphics[width=\linewidth]{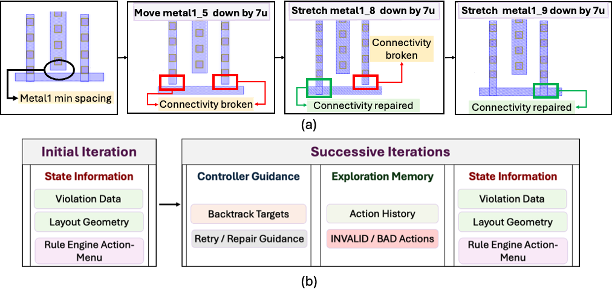}
  \caption{a) Connectivity Engine actions b) Dynamic Prompt Assembly from initial to successive iterations, incorporating exploration history and controller guidance}
  \label{fig:prompt_conn}
  \Description {The figure shows how the prompt given to the LLM-agent changes dynamically. Initially we begin with only the violation data, layout geometry and the rule-engine action-menu. But as the exploration continues, backtrack targets, previously tried actions, their efficacy is added on to the prompt, giving the LLM historical context of the exploration.}
\end{figure}

The RAT is the core decision-making element of the framework. It utilizes the reasoning ability of an LLM to navigate the layout repair search space. 
Instead of relying on unguided stochastic search, we incorporate a two-pass reasoning system. In the first pass, we inject dynamic state-aware context at runtime (action history, state memory), and a discrete menu of candidate actions generated by the Rule Engine (refer to Sec.~\ref{subsubsec:rule-eng}). This allows the LLM agent to perform semantic evaluation and generate informed responses based on spatial and historical context. In the second pass, these semantically reasoned outputs are parsed to generate tool-calls to progress the search. Fig.~\ref{fig:prompt_conn}(a) illustrates how the prompt evolves dynamically to capture this expanding state-aware context. The agent navigates the search space with five tools: MOVE, SHRINK, and EXPAND modify a shape's position or dimensions along any of the four directions (top, bottom, left, right); BACKTRACK returns to a previous node when the current branch exhausts its budget or fails to improve; and STOP terminates exploration when residual violations resist extensive search or the budget is exhausted.

\begin{algorithm}[b]
\caption{Rule Engine Candidate Generation}
\label{alg:rule-engine}
\begin{algorithmic}[1]
\Require Violations $\mathcal{V}$, layout geometry $\mathcal{G}$, ruledeck $\mathcal{R}$, action history $\mathcal{H}$
\Ensure Structured candidate menu $\mathcal{A}$
\State $\mathcal{A} \gets \emptyset$
\For{each violation $v_i \in \mathcal{V}$}
    \State Read rule category, layers, threshold $t_i$ from $\mathcal{R}$
    \State Construct hypotheses $\mathcal{P}_i$ from nearby shapes on
    rule-relevant layers
    \State Rank $\mathcal{P}_i$ by proximity and consistency; retain
    best matches
    \For{each retained hypothesis $h \in \mathcal{P}_i$}
        \State Measure rule deficit $\delta(h)$ against $t_i$
        \If{spacing} generate separation moves, facing-edge shrinks
        \ElsIf{width} generate deficient-edge expansions
        \ElsIf{enclosure} generate outer expansions, inner adjustments
        \ElsIf{overlap} generate separating moves or edge shrinks
        \EndIf
        \State Snap magnitudes to manufacturing grid
        \State Classify as legal / risk-annotated / blocked via
        editability, boundary, min-width, and action history ($\mathcal{H}$) checks
        \State Add legal and risk-annotated candidates to $\mathcal{A}$
    \EndFor
\EndFor
\State Merge equivalent candidates across markers; \Return $\mathcal{A}$
\end{algorithmic}
\end{algorithm}

\subsubsection{\textbf{Rule Engine}} \label{subsubsec:rule-eng}
If left unconstrained, the potential search space (S) for layout repair is combinatorially explosive for an LLM agent to navigate, requiring it to simultaneously select a candidate shape, edit type, direction, and magnitude:
\[
\begin{gathered}
\mathcal{S}
= C_s \times E_{\mathrm{tp}} \times E_{\mathrm{dir}} \times E_{\mathrm{mag}}, \\
\begin{array}{@{}l@{\;}l@{\qquad}l@{}}
\multicolumn{3}{@{}l}{\text{where,}} \\
C_s
& \subseteq \mathcal{C}_{\mathrm{layout}},
& \text{is the set of candidate shapes} \\

E_{\mathrm{tp}}
& \subseteq \{\mathrm{move}, \mathrm{shrink}, \mathrm{expand}\},
& \text{is the set of edit types} \\

E_{\mathrm{dir}}
& \subseteq \{\mathrm{right}, \mathrm{left}, \mathrm{top}, \mathrm{bottom}\},
& \text{is the set of edit directions} \\

E_{\mathrm{mag}} & \subseteq \{k \cdot \Delta_{\mathrm{grid}} \mid k \in \mathbb{Z}^+\}, & \text{is the set of edit magnitudes}

\end{array}
\end{gathered}
\]

To enable effective exploration, we introduce the Rule Engine, which discretizes the vast search space into a bounded set of valid geometric transformations, generating a library of potential fixes for each violation. A Calibre marker does not always uniquely identify the offending geometry; the engine therefore constructs rule-specific geometric hypotheses from shapes near the marker (facing pairs for spacing, deficient shapes for minimum width, inner--outer families for enclosure and containment, and intersecting pairs for overlap) and ranks them by marker proximity and geometric consistency. Algorithm~\ref{alg:rule-engine} summarizes the overall flow: for each retained hypothesis, candidates are generated and classified as \textit{legal}, \textit{risk-annotated} (selectable, with flagged collateral risk), or \textit{blocked}. Rule thresholds and layer relations are read from the ruledeck, and the generated actions span the local spacing, minimum-width, enclosure, extension, containment, and overlap checks evaluated in this work (the engine does not generate other topology-changing actions). While each individual action can resolve the targeted violation in isolation, it may inadvertently introduce new violations in the context of the layout. Consequently, the task of the LLM agent shifts from raw action-generation to strategic action-selection and trade-off evaluation.

\subsubsection{\textbf{Connectivity Engine}} \label{subsubsec:conn-eng}
This engine acts as a post-processor to the agent’s raw output to ensure the proposed actions do not break essential connectivity established at the root node. When the agent chooses an action, this engine performs a local connectivity check. If connectivity is preserved, the raw output is passed on to the MC for later stages. However, if severed, the engine \textit{bundles} the agent’s output with a compensatory repair sequence to fix the broken connection. This \textit{bundle} is passed on to the MC, which then enforces the LMAT operation. Compensatory bundles are not trusted: they pass through the same Calibre DRC/LVS verification as any other edit and are backtracked identically if they introduce violations or fail LVS. Fig. \ref{fig:prompt_conn}(a) illustrates how the connectivity repair is performed.

\subsection{Layout Management Toolkit (LMAT)} \label{subsec:LMAT}
The LMAT is the framework’s physical execution layer, translating RAT’s semantic decisions into physical layout modifications. To achieve real-time execution, 
we implement a Python-to-SKILL bridge that enables programmatic layout manipulations within \textit{Cadence Virtuoso}. The MC transmits a parsed, SKILL-compliant instruction set, comprising discrete or bundled repairs, to the LMAT, which executes these edits to generate the new physical node by implementing the RAT's proposed action. 

\subsubsection{\textbf{Multi-stage verification}}
The LMAT performs verification checks to analyze the edit's impact on the layout's physical and topological integrity. To maintain industry-standard sign-off accuracy, LMAT uses \textit{Siemens Calibre (nmDRC and nmLVS)} as the primary verification engine.

\paragraph{\textbf{DRC evaluation}:} \label{DRC}
Every generated node undergoes a mandatory DRC evaluation (using \textit{Calibre nmDRC}) to either verify whether all violations were resolved or extract the updated violation data required to drive subsequent iterations of the search.

\paragraph{\textbf{LVS Health Checks}:} \label{LVS}
Although autonomous LVS repair is beyond the scope of this work, LVS is employed as an electrical-topology guardrail. A mandatory check (using \textit{Calibre nmLVS}) is triggered for DRC-clean states, and for intermediate states where an edit introduces a new overlap between distinct polygons. If Calibre reports a mismatch relative to the root (device- or pin-count discrepancy, unintended short, or an open connection), the state is discarded and search continues for a better solution. This ensures that unintended topological violations are immediately caught.

\subsubsection{\textbf{Post-evaluation backtrack \& MC feedback}}
The LMAT acts as the feedback mechanism for the MC’s search strategy. If any generated node fails the mandated LVS health check, it is flagged invalid. This signals an immediate backtrack to the MC, which then prunes the non-functional branch, ensuring the search continues until a viable solution is found. A \textit{strict success} therefore denotes a DRC-clean layout whose extracted topology remains LVS-equivalent to the original cell.

\subsection{Global Memory} \label{subsec:memory}
To refine the search in real-time, MC maintains a global memory bank that serves as a comprehensive index of all exploration artifacts. 
It tracks individual shape modifications to ensure layout-geometry consistency. Additionally, it maintains a state memory and node fingerprint database, logging visited states, executed actions and their respective efficacy.
Actions already executed at a node, or flagged as \textit{bad} or \textit{invalid}, are recorded in a localized per-node memory. During the reasoning phase, this historical context is dynamically injected into the RAT’s prompt (Fig.~\ref{fig:prompt_conn}(a)). Consequently, if the RAT attempts to propose such an action, it is forced to re-evaluate and select an alternative. This dynamic feedback loop effectively prevents oscillation between previously failed states.

\section{Results} \label{sec:results}
\subsection{Experimental Setup} \label{subsec:exp_setup}

\paragraph{\textbf{Evaluation benchmark:}}

A public corpus of DRC-repair benchmark does not currently exist. Therefore, we evaluate DRC-Aid on a 60-layout test suite derived from DRC-clean FreePDK45 standard cells~\cite{FreePDK}, constructed to span the rule families and compound-violation structure encountered in layout editing. Source polygons were decomposed into Manhattan rectangles, and eligible shapes and layers were perturbed by randomized amounts calculated relative to the ruledeck thresholds. To avoid a collection of independent single-error examples, additional violations were introduced in the neighborhoods of existing violations, producing interacting, compound cases. The resulting Calibre-reported markers define the test suite. DRC-Aid receives only the perturbed layout and Calibre-reported markers; it is given no knowledge of the injected operation, magnitude, seed, or original geometry, and its inputs match what an engineer sees mid-flow. 

As summarized in Table~\ref{tab:benchmark}, the test suite contains 285 initial markers spanning 26 checks across 8 layer families (n/p well, n/p implant, active, poly, contact, Metal1, Via1, Metal2) and is predominantly compound: 53 of 60 layouts contain multiple markers, 48 multiple rule checks, and 39 violations from multiple layer families. Repairs are thus performed in the presence of surrounding geometry and other violations, where a locally plausible edit may affect subsequent decisions. Only 5 of 60 layouts (8.3\%) are cleared by a single root-state Rule-Engine action; the rest require multi-step correction and reasoning over interacting repairs, with 45–48 of 60 layouts requiring at least one backtrack across model–seed runs.

\begin{table}[t]
\centering
\caption{Benchmark characterization.}
\label{tab:benchmark}
\footnotesize
\setlength{\tabcolsep}{3.5pt}
\renewcommand{\arraystretch}{1.06}
\begin{tabular}{@{}l c l c@{}}
\toprule
\textbf{Scale} & \textbf{Value} & \textbf{Complexity} & \textbf{Value} \\
\midrule
Layouts            & 60  & Multi-marker      & 53 (88.3\%) \\
DRC markers        & 285 & Multi-rule        & 48 (80.0\%) \\
Rule checks        & 26  & Cross-family      & 39 (65.0\%) \\
Layer families     & 8   & $\geq 6$ markers  & 19 (31.6\%) \\
\midrule
\textbf{Markers/layout} & \textbf{Layouts (\%)} & \textbf{Marker category} & \textbf{Markers (\%)} \\
\midrule
1--2 & 30.0   & Spacing (same/cross layer) & 56.8 (37.2/19.6) \\
3--5 & 38.3 & Enclosure & 30.5\\
6--8 & 15.0   & Minimum width  &  7.8   \\
9--16 & 16.7  & Overlap    & 4.9   \\
\bottomrule
\end{tabular}
\parbox{0.96\columnwidth}{\footnotesize
\textit{Note:} Layouts \% over 60 layouts; markers \% over 285 initial markers. } 
\end{table}

\paragraph{\textbf{Models and execution:}}
We deployed \textsf{DRC-Aid} on a single \\NVIDIA H200 GPU using two off-the-shelf LLMs, Qwen3-Next-80B-A3B-Instruct~\cite{qwen3} and Gemma-4-26B-A4B-it~\cite{gemma4} (hereafter Qwen-80B and Gemma-26B), served through vLLM~\cite{vllm} with 32k context (Qwen-80B used bitsandbytes quantization), top-$p$ 0.95; temperature starts at 0.2 and is raised  stepwise to at most 0.6 when the agent fails to select a valid action. Per-node branch budgets decay with search depth
under a global cap of 300 productive cycles. Neither model is fine-tuned. To capture model variation and sampling non-determinism, each model was evaluated on all 60 layouts across three independent seeds (360 model--layout--seed evaluations). All selection policies and ablations use the same layouts, search budget, and verification flow. As the implementation of \cite{Simm-Ann} is not publicly available, we reimplement its improved simulated annealing (adaptive temperature decay, Metropolis acceptance, conflict-guided neighborhood proposal, and best-solution restoration) over our Rule Engine action-menu, with the violation count as objective and the same verification flow and budget as all other policies. Connectivity Engine is disabled to match the scope of \cite{Simm-Ann}.

\subsection{Evaluation} \label{subsec:results}

We classify post-repair states into\\ $\mathcal{S}_{\mathrm{strict}}$ (DRC-clean and LVS-equivalent to root, Sec.~\ref{LVS}), \\$\mathcal{S}_{\mathrm{pareto}}$ (DRC-reduced and LVS-equivalent), and \\$\mathcal{S}_{\mathrm{drc}}$ (DRC-clean regardless of LVS, tracked for ablations). 

Since every run achieved at least a partial LVS-safe reduction, $\mathcal{S}_{\mathrm{strict}} \cup \mathcal{S}_{\mathrm{pareto}} =
\mathcal{T}$, the full test suite. 

We report the \textbf{Strict Success Rate} $\mathrm{SSR} = |\mathcal{S}_{\mathrm{strict}}|/|\mathcal{T}|$, 
\\\textbf{Pareto Success Rate} $\mathrm{PSR} = |\mathcal{S}_{\mathrm{pareto}}|/|\mathcal{T}|$, 
and for ablations, $\mathrm{DCO} = |\mathcal{S}_{\mathrm{drc}}|/|\mathcal{T}|$. 

Violation clearance over a subset $\mathcal{X}$ is \\$\mathrm{VR}(\mathcal{X}) = \sum_{i \in \mathcal{X}} (V_i^{\mathrm{init}} - V_i^{\mathrm{rem}}) / \sum_{i \in \mathcal{X}} V_i^{\mathrm{init}}$, from which we report the \textbf{Total Violation Reduction} $\mathrm{TVR} = \mathrm{VR}(\mathcal{T})$ and the \textbf{Partial Violation Reduction} $\mathrm{PVR} = \mathrm{VR}(\mathcal{S}_{\mathrm{pareto}})$.

\begin{table}[t]
\centering
\caption{Seed-wise framework stability. (All values in \%)}
\label{tab:main_results}
\small
\setlength{\tabcolsep}{4pt}
\renewcommand{\arraystretch}{1.08}
\begin{tabular}{@{}l l c c c c@{}}
\toprule
\textbf{Model} & \textbf{Metric} & \textbf{Seed A} & \textbf{Seed B} & \textbf{Seed C} & \textbf{Mean $\pm$ std} \\
\midrule
\multirow{3}{*}{Gemma-26B}
  & SSR & 96.7 & 86.7 & 93.3 & $92.2 \pm 5.1$ \\
  & TVR & 99.6 & 97.2 & 97.9 & $98.2 \pm 1.3$ \\
  & PVR & 83.3 & 85.0 & 88.8 & $85.7 \pm 2.8$ \\
\midrule
\multirow{3}{*}{Qwen-80B}
  & SSR & 91.7 & 93.3 & 93.3 & $92.8 \pm 1.0$ \\
  & TVR & 98.6 & 97.8 & 98.9 & $98.4 \pm 0.6$ \\
  & PVR & 80.0 & 74.1 & 78.6 & $77.6 \pm 3.1$ \\
\bottomrule
\end{tabular}
\begin{minipage}{\linewidth}
\footnotesize
\raggedright
\textit{Note:} SSR over the 60-layout suite (PSR $= 100\% - \mathrm{SSR}$); TVR over aggregate initial violations; PVR within Pareto cases. 
\end{minipage}
\end{table}

As shown in Table~\ref{tab:main_results}, DRC-Aid reaches $\sim$92\% SSR across both models with $\sim$98\% TVR, versus 1.6–6.6\% SSR for static prompting (Table \ref{tab:ablation_results}). The $\sim$7.5\% best-effort cases still reduce violations by 78–86\% (PVR) while preserving LVS-equivalence, returning a verified partial repair rather than a failure. All edits were made within the layout boundary, thus additional area cost was not incurred.
Every layout in the suite was strictly resolved by at least two of the six model–seed runs, indicating that residual failures reflect sampling non-determinism and the finite per-node budget rather than intrinsic unsolvability. Failed runs are distributed across the difficulty spectrum, including on layouts that other seeds resolve cleanly.
Conditional on strict success, repairs take a median 8–9 Calibre nmDRC and 2 nmLVS calls, 4.4 min (Gemma-26B) / 11.4 min (Qwen-80B) (Table~\ref{tab:policy_cost}), and 40–44k input / 2.5–3.4k output tokens per repair (input tokens include the injected layout-geometry context (serialized shape coordinates, violation metadata, and Rule Engine action-menus) that grows with search depth). As reflected by comparing LLM and non-LLM policies at matched verification workloads, Gemma-26B's inference overhead is negligible relative to EDA-tool variance; Qwen-80B adds ~25–30 s per LLM generation, accounting for roughly half its wall-clock time. 

\begin{table*}[t]
\centering
\small
\caption{Ablation Study of Framework Components Across Evaluation Cases (All values in \%).}
\label{tab:ablation_results}
\begin{tabular}{p{0.1\linewidth} cccccc c >{\columncolor{gray!15}}c c c >{\columncolor{gray!15}}c c}
\toprule
& \multicolumn{6}{c}{\textbf{Components Enabled}} & \multicolumn{3}{c}{\textbf{Gemma 26B}} & \multicolumn{3}{c}{\textbf{Qwen3 80B}} \\
\cmidrule(lr){2-7} \cmidrule(lr){8-10} \cmidrule(lr){11-13}
\multirow{2}{=}{\textbf{Configuration}} 
& \textbf{2-Pass} & \textbf{DRC}  & \textbf{Iter.}   & \textbf{Mem.}   & \textbf{Rule} & \textbf{LVS}   & & & & & & \\
& \textbf{Reason} & \textbf{F.B.} & \textbf{Search} & \textbf{Prune} & \textbf{Eng.} & \textbf{Check} & \multirow{-2}{*}{\textbf{TVR}} & \multirow{-2}{*}{\textbf{SSR}} & \multirow{-2}{*}{\textbf{DCO}} & \multirow{-2}{*}{\textbf{TVR}} & \multirow{-2}{*}{\textbf{SSR}} & \multirow{-2}{*}{\textbf{DCO}} \\
\midrule
\multirow{2}{=}{LLM Prompting} 
& -- & \checkmark & -- & -- & -- & -- & 8.0 & 3.3 & 3.3 & 8.9 & 0 & 1.66 \\
& \checkmark & \checkmark & -- & -- & -- & -- & 21.0 & 6.6 & 6.6 & 10.6 & 1.66 & 1.66 \\
\midrule
\multirow{2}{=}{DRC Iteration} 
& -- & \checkmark & \checkmark & -- & -- & -- & 17.7 & 5.0 & 11.6 & 21.2 & 6.7 & 21.7 \\
& \checkmark & \checkmark & \checkmark & -- & -- & -- & 62.7 & 30.0 & 46.6 & 85.8 & 18.3 & 81.7 \\
\midrule
\multirow{2}{=}{Memory \& Pruning} 
& -- & \checkmark & \checkmark & \checkmark & -- & -- & 19.8 & 20.0 & 31.6 & 23.0 & 11.7 & 35.0 \\
& \checkmark & \checkmark & \checkmark & \checkmark & -- & -- & 69.1 & 35.0 & 68.3 & 67.7 & 23.3 & 71.7 \\
\midrule
DRC-Aid (no LVS)
& \checkmark & \checkmark & \checkmark & \checkmark & \checkmark & -- & 98.2 & 73.8 & 97.8 & 98.4 & 74.5 & 95.0 \\
\midrule
\textbf{DRC-Aid (Full)} 
& \checkmark & \checkmark & \checkmark & \checkmark & \checkmark & \checkmark & \textbf{98.2} & \textbf{92.2} & \textbf{97.8} & \textbf{98.4} & \textbf{92.8} & \textbf{95.0} \\
\bottomrule
\end{tabular}
\smallskip
\begin{minipage}{0.95\textwidth}
\centering
    \emph{\textbf{Component Legend}:}\quad \textbf{2-Pass}: 2-Pass Reasoning (enabled-vs-disabled) \quad \textbf{DRC F.B.}: DRC Feedback \quad \textbf{Iter. Search}: Iterative Search \quad \textbf{Mem. Prune}: Memory \& Pruning \quad \textbf{Rule Eng.}: Rule Engine \quad \textbf{LVS Check}: LVS-incorporated iterations
  \end{minipage}
\end{table*}

\begin{table*}[t]
\centering
\footnotesize
\caption{Action-selection policy ablation}
\label{tab:policy_cost}
\begin{tabular*}{\textwidth}{
@{\extracolsep{\fill}}
l l c c c c c c c
@{}}
\toprule
\textbf{Setting} &
\textbf{Policy} &
\multicolumn{3}{c}{\textbf{SSR (\%)}} &
\textbf{TVR (\%)} &
\textbf{DCO (\%)} &
\textbf{Time (mins)} &
\textbf{DRC/LVS (calls)} \\
\cmidrule(lr){3-5}

& &
\textbf{All} &
\textbf{$\geq6$ markers} &
\textbf{$<6$ markers} &
& & & \\

\midrule

\multirow{3}{*}{Shared menu}
& Random
& $54.4\pm8.2$
& $33.3\pm11.0$
& $64.2\pm7.0$
& $89.5\pm3.6$
& $72.8\pm3.5$
& 5.0 [2.0, 11.7]
& 7 [3, 17.5] / 3 [2, 7] \\

& Deterministic
& $83.3$
& $63.2$
& $92.7$
& $95.1$
& $88.3$
& 3.3 [2.0, 5.8]
& 5 [3, 8.3] / 1 [1, 2] \\

& Softmax ($\tau=0.5$)
& $82.8\pm2.5$
& $63.2\pm5.3$
& $91.9\pm1.4$
& $91.5\pm5.4$
& $87.8\pm1.9$
& 3.4 [2.0, 7.3]
& 5.5 [3, 12.8] / 1 [1, 2] \\
\midrule

Prior work
& SA-RE
& $38.3\pm6.0$
& $24.6\pm6.1$
& $44.7\pm6.1$
& $86.0\pm0.0$
& $63.9\pm2.5$
& 4.3 [1.8, 10.5]
& 8 [2.3, 19.8] / 3 [2, 6] \\
\midrule

\multirow{2}{*}{DRC-Aid}
& Gemma-26B
& $92.2\pm5.1$
& $82.5\pm8.0$
& $96.7\pm3.7$
& $98.2\pm1.3$
& $97.8 \pm 2.5$
& 4.4 [1.8, 9.6]
& 8 [3, 17] / 2 [1, 5] \\

& Qwen-80B
& $92.8\pm1.0$
& $84.2\pm5.3$
& $96.7\pm1.4$
& $98.4\pm0.6$
& $95.0 \pm 0.0$
& 11.4 [4.0, 30.9]
& 9 [3, 19.8] / 2 [1, 5] \\

\bottomrule
\end{tabular*}
\begin{minipage}{\textwidth}
\footnotesize
\raggedright
\textit{Note:}
SSR/TVR/DCO are mean~$\pm$~std over three seeds, except for the seed-independent deterministic policy. Costs are median [Q1,~Q3] over strict-success runs. All policies use the same Rule-Engine candidate menu, state budget, and Calibre verification tools. SA-RE omits connectivity-aware bundling to follow the scope of \cite{Simm-Ann}; its comparison therefore evaluates the prior-work-style system rather than isolating action selection alone. Cost statistics are conditional on SSR runs; thus
not unconditional efficiency comparisons.
\end{minipage}
\end{table*}

\subsection{Ablation Studies} \label{subsec:ablation}

\paragraph{\textbf{Component ablation:}}
Table~\ref{tab:ablation_results} quantifies each module's impact along TVR, DCO, and SSR. Across all configurations, \textbf{2-pass reasoning} consistently outperforms direct 1-pass tool-call generation, but the static prompting baseline fails regardless (SSR $\leq$ 6.6\%). Adding \textbf{DRC-feedback iteration} reveals raw geometric capability (Qwen: 85.8\% TVR, 81.7\% DCO), but blind edits degrade topology (18.3\% SSR). \textbf{Memory and pruning} prevents cyclic oscillation and trades blind clearance for topological integrity (Qwen DCO drops to 71.6\% while SSR rises to 23.3\%), yet the unconstrained action space remains too vast. Coupling the LLM with the \textbf{Rule Engine} lifts TVR above 98\% and DCO to 95--97.8\%, but geometric heuristics alone cap SSR at ${\sim}74\%$. Finally, \textbf{LVS-incorporated iteration} preserves TVR/DCO (as they are LVS-independent metrics) while ensuring topological integrity, surging SSR to ${\sim}92\%$ for both models. Automated repair thus requires the structured coupling of constrained spatial reasoning with multi-domain (DRC and LVS) verification --- no single component suffices.

\paragraph{\textbf{Selection-policy and prior-work baselines:}} To isolate the contribution of LLM selection from the Rule Engine, we compare three non-LLM selectors and a reimplementation of the simulated-annealing repair of \cite{Simm-Ann} under identical candidate generation, verification flow, search, and budget, 
The deterministic heuristic selects the top-ranked action under a fixed lexicographic priority over the same pruned menu: (i) violation-group coverage, (ii) primary over conflict-escape class, (iii) smaller edit magnitude, (iv) deterministic tie-break. The softmax policy samples over the identical heuristic ranking with $\tau$=0.5; $\tau\rightarrow$0 recovers the deterministic ranker and $\tau\rightarrow\infty$ approaches uniform. This isolates whether the LLM's advantage is merely sampling diversity over the same ranking.

Random selection solves 54.4\%, so candidate generation alone does not trivialize the task; the deterministic heuristic reaches 83.3\% at the lowest runtime. The LLM policies reach 92.2–92.8\%, and the margin widens on $\geq$6-marker layouts: 63.2\% (heuristic) vs. 82.5–84.2\% (LLM). The softmax policy does not materially improve over the deterministic ranker, indicating that the observed LLM gain is not explained by simple stochastic sampling over this fixed ranking. The result is consistent with the LLM making use of state, geometry, and action-history context. The SA baseline behaves differently in kind. Because \cite{Simm-Ann} optimizes DRC clearance without LVS-aware acceptance, DCO matches its scope; SA-RE reaches 63.9\% DCO on our suite, compared with 95.0–97.6\% for DRC-Aid, and 38.3\% SSR once LVS-equivalence is required.

\section{Conclusion} \label{sec:conc}
In this work, we presented \textbf{DRC-Aid}, a closed-loop agentic framework that automates local DRC repair by coupling a deterministic Rule Engine, budgeted DFS exploration with global memory, and LLM-based action selection with verification-in-the-loop. Evaluations demonstrate that DRC-Aid achieves $\sim$98\% total violation reduction, with $\sim$92.5\% of cases reaching complete resolution, a substantial improvement over static prompting, which peaks at \textbf{6.6\%} complete resolution. Under identical evaluation infrastructure, random selection yielded 54.4\% completion, and a deterministic ranker yielded 83.3\%, with the LLM's advantage widening on multi-violation layouts. 

The present evaluation verifies geometric legality and topology preservation on conventional geometric DRCs in FreePDK45. Natural next steps include extending the verification loop to parasitic- and timing-aware acceptance criteria, broadening rule-deck coverage across technologies, and enriching the action space with repairs for congested post-route scenarios. More broadly, DRC-Aid demonstrates that bounded, frequently repeated correction loops can be automated by constrained, tool-grounded agents.

\begin{acks}
This work was supported in part by the Center for the Co-Design of Cognitive Systems (CoCoSYS), a research center under the Joint University Microelectronics Program (JUMP) 2.0, a Semiconductor Research Corporation (SRC) initiative sponsored by DARPA, and by National Science Foundation.
\end{acks}

\bibliographystyle{ACM-Reference-Format}
\bibliography{sections/ref}

\end{document}